\documentclass{article}
\usepackage{spconf,amsmath,graphicx,hyperref}
\usepackage{url}
\usepackage{pifont}
\usepackage{amsfonts}
\usepackage{algorithmic}
\usepackage{algorithm}
\usepackage{array}
\usepackage{textcomp}
\usepackage{stfloats}
\usepackage{url}
\usepackage{verbatim}
\usepackage{cite}
\usepackage{booktabs}
\usepackage{tabularx} 
\usepackage{subcaption}  
\usepackage{caption}
\usepackage{float}
\usepackage{threeparttable}
\usepackage{multirow}
\usepackage{booktabs}
\usepackage{hyperref}
\title{mmWave-Diffusion: A Novel Framework for Respiration Sensing Using Observation-Anchored Conditional Diffusion Model}

\name{\begin{tabular}{c}
Yong Wang$^{*,1}$, Qifan Shen$^{*,1}$,Bao Zhang$^{1}$,Zijun Huang$^{1}$,Chengbo Zhu$^{1}$, Shuai Yao$^{1}$,Qisong Wu$^{\dag,1,2}$
\thanks{$^*$ The authors contributed equally. $^\dag$ Corresponding author.}
\thanks{This work was supported by the National Natural Science Foundation of China under Grants No. 12374421, and  62293552, in part by  Open fund of the Science and Technology on Underwater Test and Control Laboratory (No. 2024-CXPT-GF-JJ-043), and supported by the Fundamental Research Funds for the Central Universities under Grant No. 2242025F20003.}
\end{tabular}}
\address{$^1$Key Laboratory of Underwater Acoustic Signal Processing of Ministry of Education,\\ Southeast University, Nanjing, 211189, China\\
$^2$Purple Mountain Laboratories, Nanjing, 211111, China\\}

\begin{document}
\ninept
\maketitle
\begin{abstract}
Millimeter-wave (mmWave) radar enables contactless respiratory sensing, yet fine-grained monitoring is often degraded by nonstationary interference from body micromotions. To achieve micromotion interference removal, we propose mmWave-Diffusion, an observation-anchored conditional diffusion framework that directly models the residual between radar phase observations and the respiratory ground truth, and initializes sampling within an observation-consistent neighborhood rather than from Gaussian noise—thereby aligning the generative process with the measurement physics and reducing inference overhead. The accompanying Radar Diffusion Transformer (RDT) is explicitly conditioned on phase observations, enforces strict one-to-one temporal alignment via patch-level dual positional encodings, and injects local physical priors through banded-mask multi-head cross-attention, enabling robust denoising and interference removal in just 20 reverse steps. Evaluated on 13.25 hours of synchronized radar–respiration data, mmWave-Diffusion achieves state-of-the-art waveform reconstruction and respiratory-rate estimation with strong generalization. Code repository: \href{https://github.com/goodluckyongw/mmWave-Diffusion}{https://github.com/goodluckyongw/mmWave-Diffusion}. 
\end{abstract}
\section{Introduction}
\label{sec:intro}

Continuous, unobtrusive monitoring of respiratory patterns is critical for chronic disease management, sleep assessment, and home-based care \cite{Wang2024RF-GymCare,Zheng2021MoRe-Fi}. Owing to its contactless operation and privacy-preserving nature, millimeter-wave (mmWave) radar is a promising sensing modality for this purpose \cite{Wang2024MM-FGRM}. However, in real-world environments, fine-grained monitoring remains challenging: body micromotions induce nonstationary artifacts that perturb radar-phase measurements and obscure subtle respiratory dynamics \cite{Qiao2025Millimeter,Wang2025GAWNet}.

In radar-based respiratory monitoring, early approaches relied on conventional signal-processing techniques (e.g., filtering and mode decomposition) \cite{Zhang2023Pi}. However, the stationarity or independence assumptions underlying these methods render them highly sensitive to time-varying interference in radar returns, often limiting performance to coarse respiratory-rate estimation and precluding high-fidelity waveform reconstruction \cite{Wang2024RF-GymCare,Wang2026Fine}. Recently, deep learning models (e.g., CNNs and Transformers) have sought to learn direct mappings from noisy radar phase or I/Q signals to clean respiration waveforms \cite{Wang2024MM-FGRM,Wu2024Contactless}. Despite their high capacity, such “black-box” regressors are prone to overfitting environment-specific noise and may yield physiologically implausible outputs. Generative alternatives, such as Variational Autoencoders (VAEs), have also been explored \cite{Zheng2021MoRe-Fi,Bauder2025MM-MURE}; however, the variational objective’s simplified posterior assumptions (typically Gaussian) constrain expressiveness, preventing faithful capture of complex, multimodal distributions.

Diffusion models have recently emerged as a powerful generative paradigm, advancing image synthesis and time-series forecasting \cite{Zhang2025Diffusion,Rasul2021Autoregressive}. Their stepwise denoising mechanism enables modeling of complex, nonstationary, non-Gaussian degradations  \cite{Ho2020Denoising,Dhariwal2021Diffusion}, which naturally aligns with the challenges of radar-based respiratory monitoring. Motivated by this insight, we propose mmWave-Diffusion, an observation-anchored conditional diffusion framework for fine-grained radar respiratory sensing. To effectively remove micromotion-induced interference, the forward process progressively injects perturbations along the residual between the observed radar phase and the target respiration signal; in the reverse process, the model conditions explicitly on the phase observations and initializes the sampling trajectory within an Observation-Consistent Neighborhood (OCN). As denoising proceeds, micromotion artifacts are progressively suppressed, yielding a high-fidelity respiratory waveform. This observation anchoring avoids the lengthy noise-start sampling of standard diffusion while preserving consistency with the measurement physics.

To instantiate the framework, we design the Radar Diffusion Transformer (RDT), inspired by DiT \cite{Peebles2023Scalable}. RDT processes the main sequence (initialized from OCN to define the reverse denoising trajectory toward the clean waveform) and the conditional radar-phase sequence in parallel. Both streams are tokenized into non-overlapping patches and assigned independent positional encodings, enforcing strict one-to-one temporal correspondence. To encode locality, RDT employs banded-mask multi-head cross-attention, injecting the prior that radar echoes correlate primarily with temporally proximate respiration and thereby suppressing long-range mismatches and spurious associations. These design choices integrate radar observations throughout mmWave-Diffusion’s generative process, enabling targeted removal of micromotion interference and reconstruction of fine-grained respiratory waveforms.

\section{Method}
\label{sec:method}
\begin{figure*}[t]
\centerline{\includegraphics[width=0.96\linewidth]{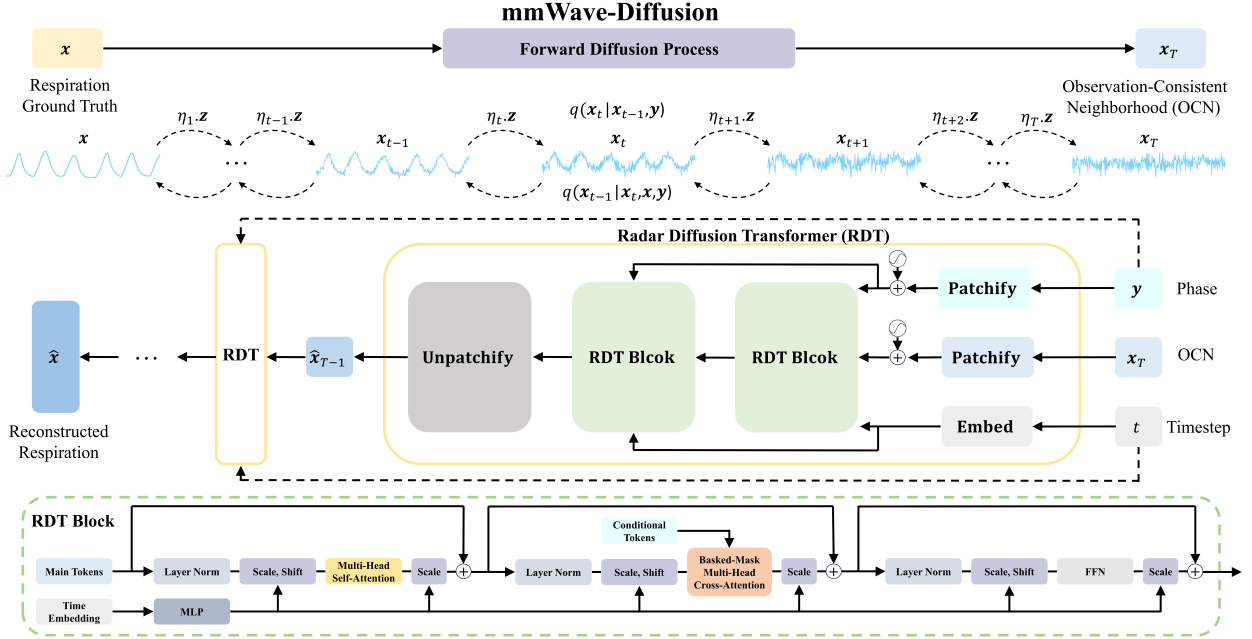}}
\caption{Overview of mmWave-Diffusion with the Radar Diffusion Transformer (RDT) architecture.}
\label{FIG:1}
\end{figure*}

This section details the mmWave-Diffusion framework (Fig.~\ref{FIG:1}) and presents the end-to-end pipeline from radar observations to respiratory waveform reconstruction.

\subsection{Signal Processing Pipeline}
\label{ssec:pipeline}

\begin{figure}[t]
\centerline{\includegraphics[width=0.98\linewidth]{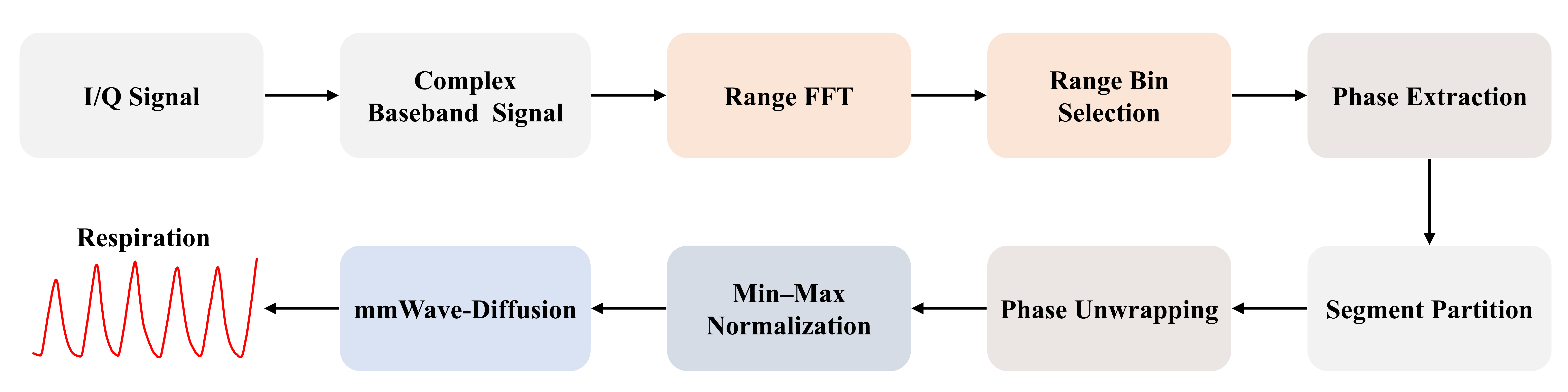}}
\caption{Signal processing pipeline.}
\label{FIG:2}
\end{figure}

Fig.~\ref{FIG:2} summarizes the end-to-end pipeline from raw radar I/Q to the reconstructed respiratory waveform. Raw I/Q streams acquired by the mmWave radar form a complex baseband signal. We first apply a range Fast Fourier Transform (FFT) to obtain a 1-D range profile; with Moving Target Indication (MTI) and a Constant False Alarm Rate (CFAR) detector, we automatically select the range bin corresponding to the subject’s chest \cite{Diao2024Review, Liaquat2024End}, and extract the phase sequence from this bin. Next, using a sliding window of \(w\) seconds with a hop of \(s\) seconds, the phase sequence is partitioned into segments of length \(l\). Each segment undergoes phase unwrapping and min–max normalization to \([0,1]\), yielding \(\mathbf{y}\in\mathbb{R}^{1\times l}\). The respiratory ground truth is segmented and normalized identically to obtain \(\mathbf{x}\in\mathbb{R}^{1\times l}\). Feeding \(\mathbf{y}\) into mmWave-Diffusion produces the reconstructed respiratory waveform \(\hat{\mathbf{x}}\in\mathbb{R}^{1\times l}\).

\subsection{mmWave-Diffusion}
\label{ssec:framework}

For micromotion-robust radar respiratory sensing, we propose mmWave-Diffusion. As illustrated in Fig.~\ref{FIG:1}, the forward process starts from the respiratory ground truth \(\mathbf{x}\) and advances along the residual path \(\mathbf{z}=\mathbf{y}-\mathbf{x}\): this residual, which encodes micromotion-induced waveform distortions, is progressively injected, driving the state toward a Gaussian with mean \(\mathbf{y}\). Conversely, the reverse diffusion process initializes within an OCN of \(\mathbf{y}\) and explicitly conditioned on \(\mathbf{y}\) at each step; together with the accompanying RDT, the observation constrains the generative trajectory, enabling iterative denoising and removal of micromotion interference, ultimately reconstructing a fine-grained respiratory waveform.

\noindent 
\textbf{Forward Process:} 
Unlike standard diffusion, whose forward process injects observation-agnostic Gaussian noise, mmWave-Diffusion progressively injects the residual $\mathbf{z}$ encoding micromotion-induced interference, thereby constructing a $T$-step Markov chain that maps the respiratory ground truth $\mathbf{x}$ to the radar observation $\mathbf{y}$. This process is governed by a monotonically increasing noise schedule $\{\eta_t\}_{t=1}^{T}$ satisfying $\eta_1 \to 0$ and $\eta_T \to 1$, where $t \in \{1,2,\ldots,T\}$ indexes the timesteps. The one-step transition kernel as follows:
\begin{equation}
\label{eq:forward-kernel}
q(\mathbf{x}_t \mid \mathbf{x}_{t-1}, \mathbf{y}) = \mathcal{N}\!\big(\mathbf{x}_t;\; \mathbf{x}_{t-1} + \alpha_t \mathbf{z},\; \kappa^2 \alpha_t \mathbf{I}\big),
\end{equation}
where $\alpha_t = \eta_t - \eta_{t-1}$ for $t>1$ and $\alpha_1=\eta_1$, $\kappa$ is a noise hyperparameter, and the chain is initialized at $\mathbf{x}_0=\mathbf{x}$. From Eq.~\eqref{eq:forward-kernel}, the marginal distribution at any timestep $t$ admits a closed form:
\begin{equation}
\label{eq:marginal}
q(\mathbf{x}_t \mid \mathbf{x}, \mathbf{y}) = \mathcal{N}\!\big(\mathbf{x}_t;\; \mathbf{x} + \eta_t \mathbf{z},\; \kappa^2 \eta_t \mathbf{I}\big).
\end{equation}
Eq.~\eqref{eq:marginal} shows that the mean of $\mathbf{x}_t$ transitions smoothly from $\mathbf{x}$ to $\mathbf{y}$ as $\eta_t$ increases, thereby completing the degradation from the respiratory signal to the radar observation. We explicitly inject a micromotion-encoded residual in the forward process, thereby preserving radar-measurement consistency throughout.

\noindent

\noindent 
\textbf{Reverse Process:} 
To effectively exploit the observations for guiding micromotion removal, we initialize the reverse diffusion within an OCN of $\mathbf{y}$,
\begin{equation}
\mathbf{x}_T = \mathbf{y} + \kappa \sqrt{\eta_T}\,\boldsymbol{\epsilon}, \qquad \boldsymbol{\epsilon}\sim\mathcal{N}(\mathbf{0},\mathbf{I}),
\end{equation}
rather than initializing from pure Gaussian noise as in standard diffusion, and then iteratively denoises to recover the respiration $\hat{\mathbf{x}}$. The reverse transition kernel $p_\theta(\mathbf{x}_{t-1} | \mathbf{x}_t, \mathbf{y})$ is learned with the RDT network parameterized by $\theta$, by minimizing the Kullback–Leibler (KL) divergence to the true posterior $q(\mathbf{x}_{t-1} | \mathbf{x}_t, \mathbf{x}, \mathbf{y})$ ~\cite{Yue2025Efficient,Liu2024Residual}. Using Eqs.~\eqref{eq:forward-kernel}--\eqref{eq:marginal}, the true posterior admits the closed form: 
\begin{equation}
\label{eq:true-posterior}
q(\mathbf{x}_{t-1}\mid \mathbf{x}_t,\mathbf{x},\mathbf{y})
= \mathcal{N}\!\Big(\mathbf{x}_{t-1};\, \tfrac{\eta_{t-1}}{\eta_t}\mathbf{x}_t + \tfrac{\alpha_t}{\eta_t}\mathbf{x},\, \kappa^2 \tfrac{\eta_{t-1}}{\eta_t}\alpha_t\,\mathbf{I} \Big).
\end{equation}
Accordingly, we parameterize $p_\theta$ as a Gaussian whose variance is fixed to match the true posterior, $\kappa^2 \tfrac{\eta_{t-1}}{\eta_t}\alpha_t\,\mathbf{I}$, and whose mean is
\begin{equation}
\label{eq:mu-theta}
\boldsymbol{\mu}_\theta(\mathbf{x}_t,\mathbf{y},t)
= \tfrac{\eta_{t-1}}{\eta_t}\mathbf{x}_t + \tfrac{\alpha_t}{\eta_t}\,\mathrm{RDT}_\theta(\mathbf{x}_t,\mathbf{y},t),
\end{equation}
where the denoiser $\mathrm{RDT}_\theta(\cdot)$ predicts the clean respiratory signal from the noisy state $\mathbf{x}_t$ conditioned on $\mathbf{y}$, progressively removing micromotion interference. Following prior work \cite{Shen2025Deep}, the training objective reduces to a Mean Squared Error (MSE) loss:
\begin{equation}
\label{eq:mse}
\mathcal{L}_{t}
= \mathbb{E}_{t,\mathbf{x},\mathbf{y}}\!\big\| \mathrm{RDT}_\theta(\mathbf{x}_t,\mathbf{y},t) - \mathbf{x} \big\|^{2},
\end{equation}
which enables RDT to denoise across noise levels and reconstruct a fine-grained respiratory waveform.

\noindent
\textbf{RDT:} As the core denoiser of mmWave-Diffusion, RDT operates along the reverse trajectory, initialized at the OCN state $\mathbf{x}_T$, and conditions on the radar-phase sequence $\mathbf{y}$ at each timestep. 

To enhance denoising against micromotion interference, RDT uses dual patch-level positional encodings for the OCN-initialized main sequence $\mathbf{x}_t$ and the conditional sequence $\mathbf{y}$ to enforce strict one-to-one temporal correspondence and mitigate positional confounds, and applies a banded-mask multi-head cross-attention that limits cross-sequence interactions to a local temporal window, explicitly encoding the locality prior that radar returns correlate primarily with temporally proximate respiration and suppressing long-range mismatches and spurious associations. In addition, we adopt timestep respacing \cite{Song2020Denoising} to sample a sparse subset of the $T$-step diffusion chain, substantially reducing inference cost.

As shown in Fig.~\ref{FIG:1}, we first partition $\mathbf{x}_t$ and $\mathbf{y}$ into non-overlapping patches and add learnable positional encodings to each, producing $N$ tokens per stream. This facilitates extraction of fine-grained temporal representations and strengthens positional alignment. In addition, the timestep $t$ is mapped by an embedding layer. The resulting main tokens, conditional tokens, and timestep embedding serve as three input streams to $L$ stacked RDT blocks.

Within each block, the main-token stream first undergoes timestep-modulated normalization \cite{Peebles2023Scalable}, followed by multi-head self-attention \cite{Wang2025SelaFD,Wang2024Speech} to model intra-sequence temporal dependencies. We then employ a banded-mask multi-head cross-attention to fuse the conditional tokens with the main stream. This mask hard-codes a locality prior by permitting interactions only within a temporal window, thereby suppressing long-range mismatches and spurious associations. The banded mask is defined elementwise as
\begin{equation}
M_{ij}=
\begin{cases}
0, & |i-j|\le u,\\
-\infty, & \text{otherwise},
\end{cases}
\label{eq:bandmask_element_en}
\end{equation}
where $i,j\in\{1,\ldots,N\}$ index the main and conditional token sequences, respectively; $u$ is the local window radius; and $\mathbf{M}=[M_{ij}]_{i,j=1}^{N}\in\mathbb{R}^{N\times N}$ is the mask matrix. For a single head, cross-attention is computed as
\begin{equation}
\mathrm{Attention}(\mathbf{Q},\mathbf{K},\mathbf{V})
=\mathrm{softmax}\!\left(\mathbf{Q}\mathbf{K}'/\sqrt{d}+\mathbf{M}\right)\mathbf{V},
\label{eq:cross_attn_en}
\end{equation}
where $\mathbf{Q}\in\mathbb{R}^{N\times d}$ is the Query from the main tokens, $\mathbf{K},\mathbf{V}\in\mathbb{R}^{N\times d}$ are the Key and Value from the conditional tokens, $d$ is the hidden dimension, $(\cdot)'$ denotes matrix transpose, and $\mathrm{softmax}(\cdot)$ is the softmax function. The cross-attention output is subsequently passed through a timestep-modulated Feed-Forward Network (FFN), whose output serves as the input to the next block. The final block’s output is inverse-patched to \(\hat{\mathbf{x}}_{t-1}\); iterating the step along the sampling path ultimately reconstructs the fine-grained waveform \(\hat{\mathbf{x}}\).

\section{Experiment}
\label{sec:experiment}

\subsection{Experimental Setup}
\label{ssec:Implementation}
To evaluate mmWave-Diffusion, we collected a synchronized radar–respiration dataset in an office environment using a 60~GHz IWR6843 FMCW radar with a DCA1000 board \cite{Lei2024Dataset}; ground-truth respiration was recorded by a GDX-RB belt \cite{Mauro2023Few-Shot}. We recorded 12 healthy subjects for 1~hour each, and written informed consent was obtained from all subjects. During acquisition, each subject sat approximately 1.2~m in front of the radar and breathed naturally. Key radar settings were: sweep bandwidth = 2~GHz; sample rate = 10~Msps; ADC samples = 512; frame periodicity = 50~ms; ramp duration = 51.2~$\mu$s. A participant-wise split is adopted, with data from 8 subjects used for training and 4 for testing~\cite{Zheng2021MoRe-Fi}. All results are reported on the test set to assess the cross-subject generalization of mmWave-Diffusion. Signals are segmented with a 20~s sliding window and a 1~s step, yielding samples of 400 points. Respiratory monitoring performance is assessed along two dimensions: waveform reconstruction (Mean Squared Error (MSE), Cosine Similarity (CS)) and frequency estimation (Mean Absolute Error (MAE), Root Mean Squared Error (RMSE), Standard Deviation (SD)).

mmWave-Diffusion was implemented in PyTorch on NVIDIA A100 Tensor Core GPUs. We trained for 100 epochs using AdamW with an initial learning rate of 0.01 and employed ReduceLROnPlateau for learning rate scheduling~\cite{Wang2025Hierarchical}. Key configuration details are: forward diffusion steps $T=1000$; reverse process in 20 steps; a noise schedule $\{\eta_t\}_{t=1}^{T}$ following Yue et al.~\cite{Yue2025Efficient} with $\eta_{1}=0.001$ and $\eta_{T}=0.999$; noise hyperparameter $\kappa=0.7$; $L=2$ RDT blocks; each input sequence partitioned into $N=20$ tokens; local window radius $u=1$; and hidden dimension $d=256$.

\begin{table}[t]
\caption{Performance of mmWave-Diffusion vs. Baselines}
\label{table:1}
\centering
\setlength{\tabcolsep}{4pt}     % 列间距
\scriptsize
\begin{tabular}{c c c c c c}
\toprule
\multirow{2}{*}{Method} &
\multicolumn{2}{c}{Waveform Reconstruction} &
\multicolumn{3}{c}{Frequency Estimation} \\
\cmidrule(lr){2-3} \cmidrule(lr){4-6}
& CS $\uparrow$ & MSE $\downarrow$ & MAE $\downarrow$ & RMSE $\downarrow$ & SD $\downarrow$ \\
\midrule
MoRe-Fi~\cite{Zheng2021MoRe-Fi}        & 0.754 & 0.104 & 1.336 & 1.835 & 1.822 \\
MM-FGRM~\cite{Wang2024MM-FGRM}         & 0.793 & 0.090 & 0.704 & 1.225 & 1.221 \\
MM-MuRe~\cite{Bauder2025MM-MURE}       & 0.760 & 0.100 & 1.949 & 2.336 & 1.416 \\
LSTM~\cite{Rasul2021Autoregressive}    & 0.798 & 0.151 & 1.058 & 1.339 & 1.211 \\
U-Net~\cite{Ho2020Denoising}    & 0.701 & 0.135 & 0.865 & 1.240 & 1.193 \\
Transformer~\cite{Peebles2023Scalable} & 0.752 & 0.108 & 1.107 & 1.428 & 1.240 \\
BPF~\cite{Islam2020NonContact}         & 0.651 & 0.254 & 7.102 & 8.655 & 5.270 \\
\textbf{mmWave-Diffusion}       & \textbf{0.811} & \textbf{0.079} & \textbf{0.631} & \textbf{1.175} & \textbf{1.035} \\
\bottomrule
\end{tabular}
\end{table}

\subsection{Performance Analysis}
\label{ssec:Performance}

To validate the effectiveness of mmWave-Diffusion, we systematically compare it against three categories of baselines across five metrics: advanced radar-based respiration monitoring methods (MoRe-Fi~\cite{Zheng2021MoRe-Fi}, MM-FGRM~\cite{Wang2024MM-FGRM}, MM-MURE~\cite{Bauder2025MM-MURE}); variants that adopt our diffusion paradigm while replacing the RDT denoiser with LSTM~\cite{Rasul2021Autoregressive}, U-Net~\cite{Ho2020Denoising}, or Transformer~\cite{Peebles2023Scalable} backbones; and a classical Band-Pass Filter (BPF) baseline~\cite{Islam2020NonContact}. As shown in Table~\ref{table:1}, mmWave-Diffusion achieves the best performance on all metrics: for waveform reconstruction, CS/MSE are 0.811/0.079; for respiratory-rate estimation, MAE/RMSE/SD are 0.631/1.175/1.035 Breaths Per Minute (BPM). These results substantiate the effectiveness of the observation-anchored diffusion framework and highlight the strength of RDT in capturing fine-grained respiratory patterns.

\begin{figure}[t]
    \centering
    \begin{subfigure}[b]{0.49\linewidth}
        \centering
        \includegraphics[width=\linewidth]{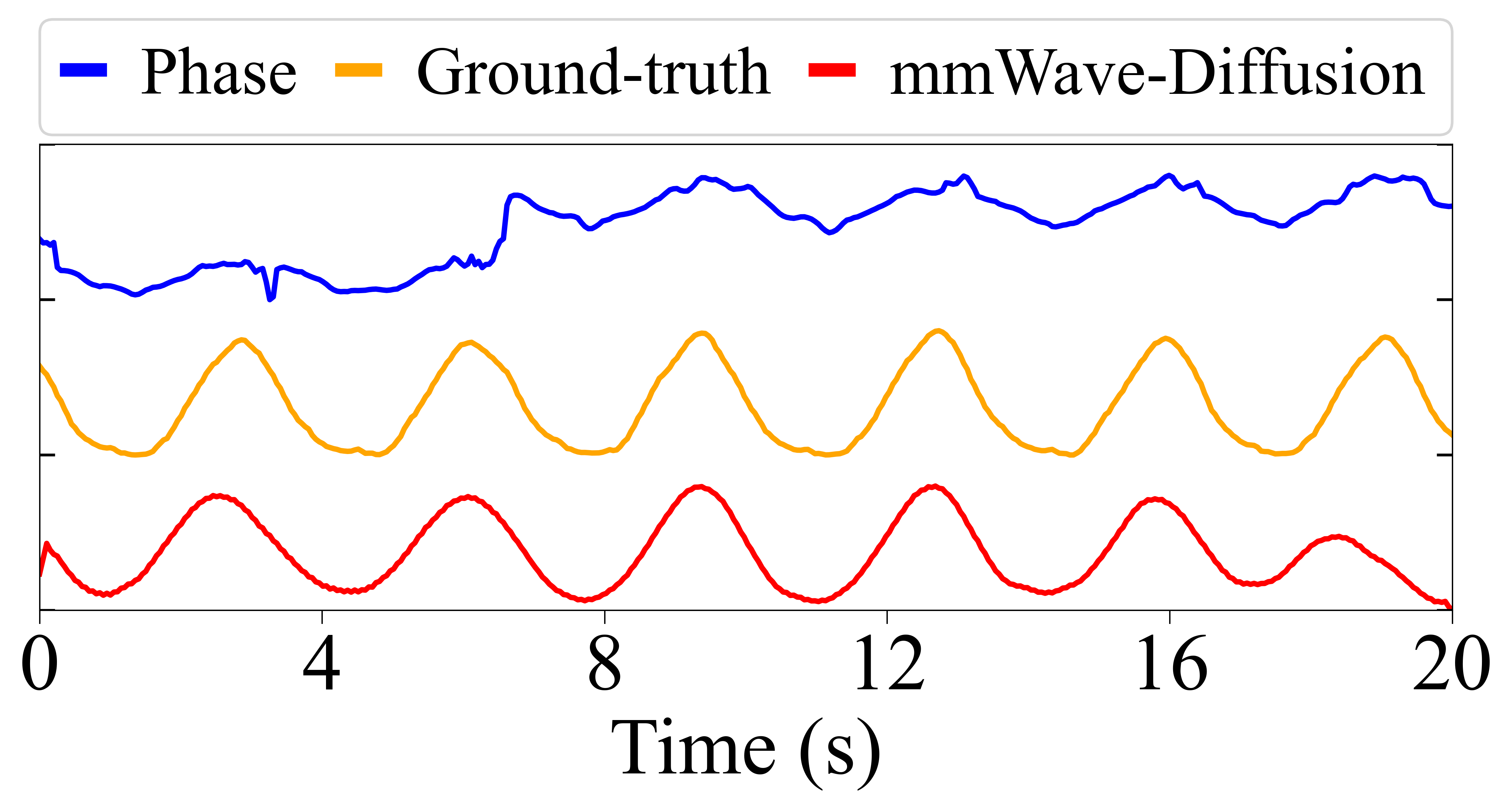}
        \caption{Smartphone use}
    \end{subfigure}
    \hfill
    \begin{subfigure}[b]{0.49\linewidth}
        \centering
        \includegraphics[width=\linewidth]{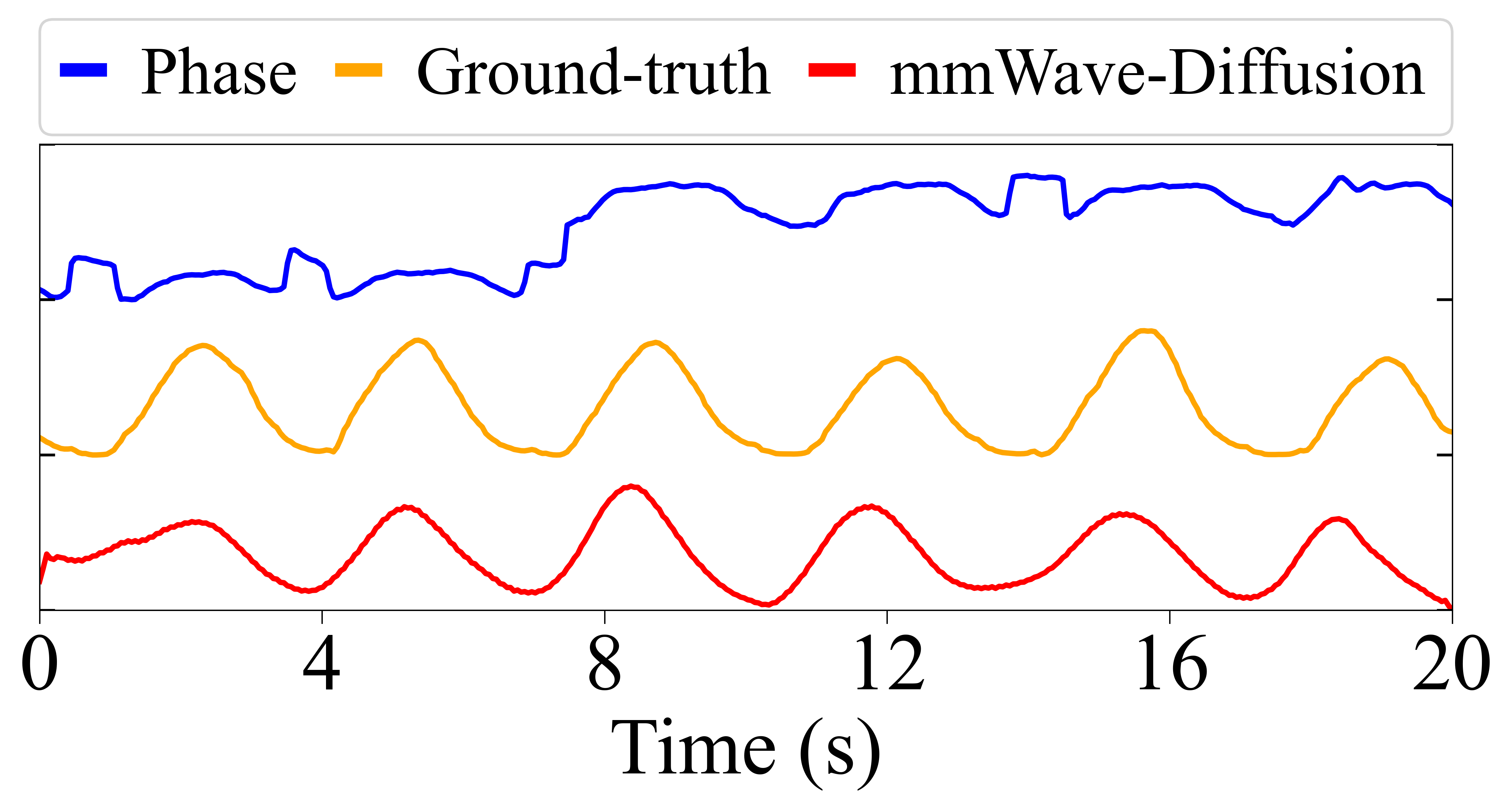}
        \caption{Conversation}
    \end{subfigure}
    \begin{subfigure}[b]{0.49\linewidth}
        \centering
        \includegraphics[width=\linewidth]{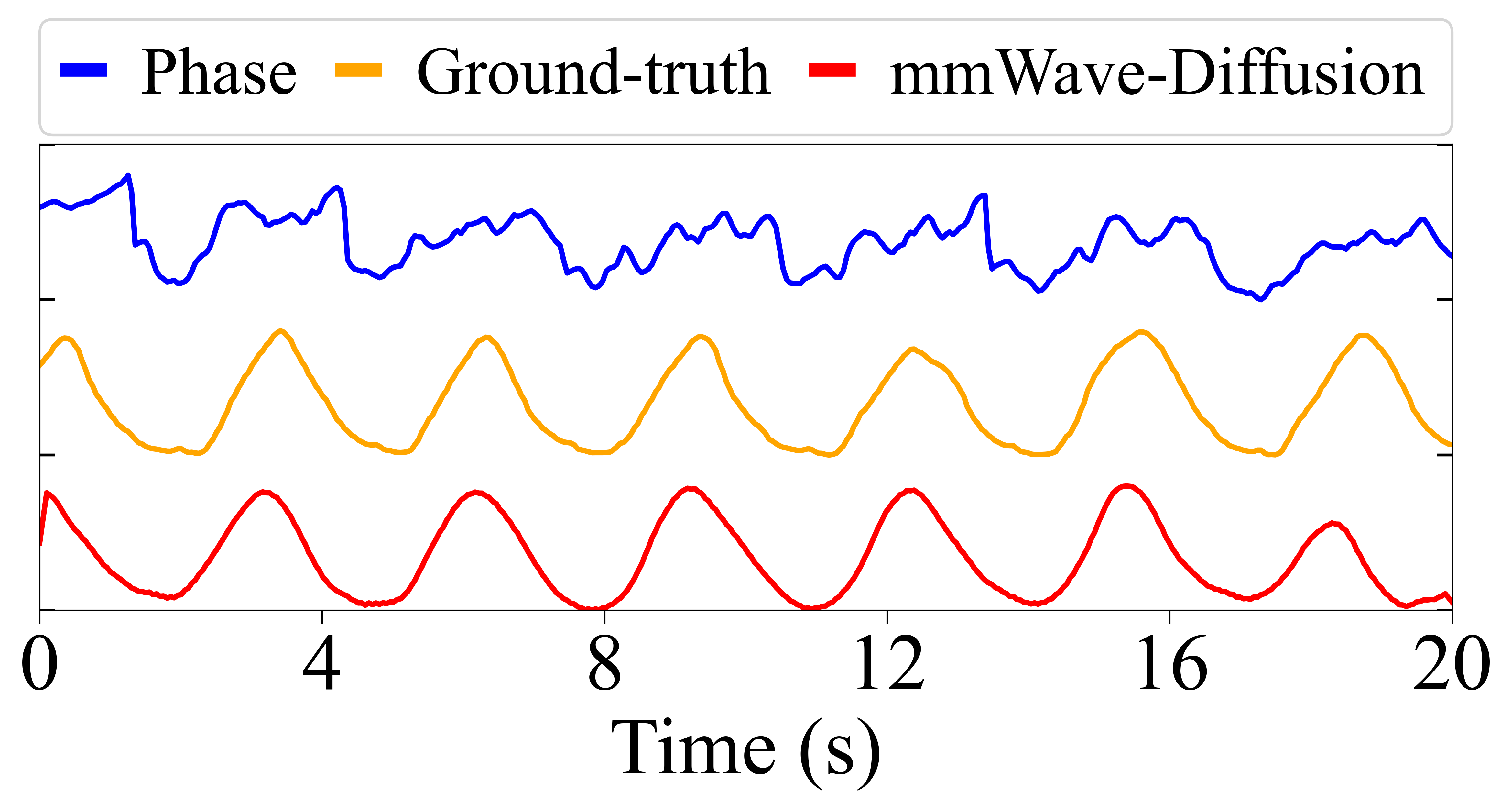}
        \caption{Leg shaking}
    \end{subfigure}
    \hfill
    \begin{subfigure}[b]{0.49\linewidth}
        \centering
        \includegraphics[width=\linewidth]{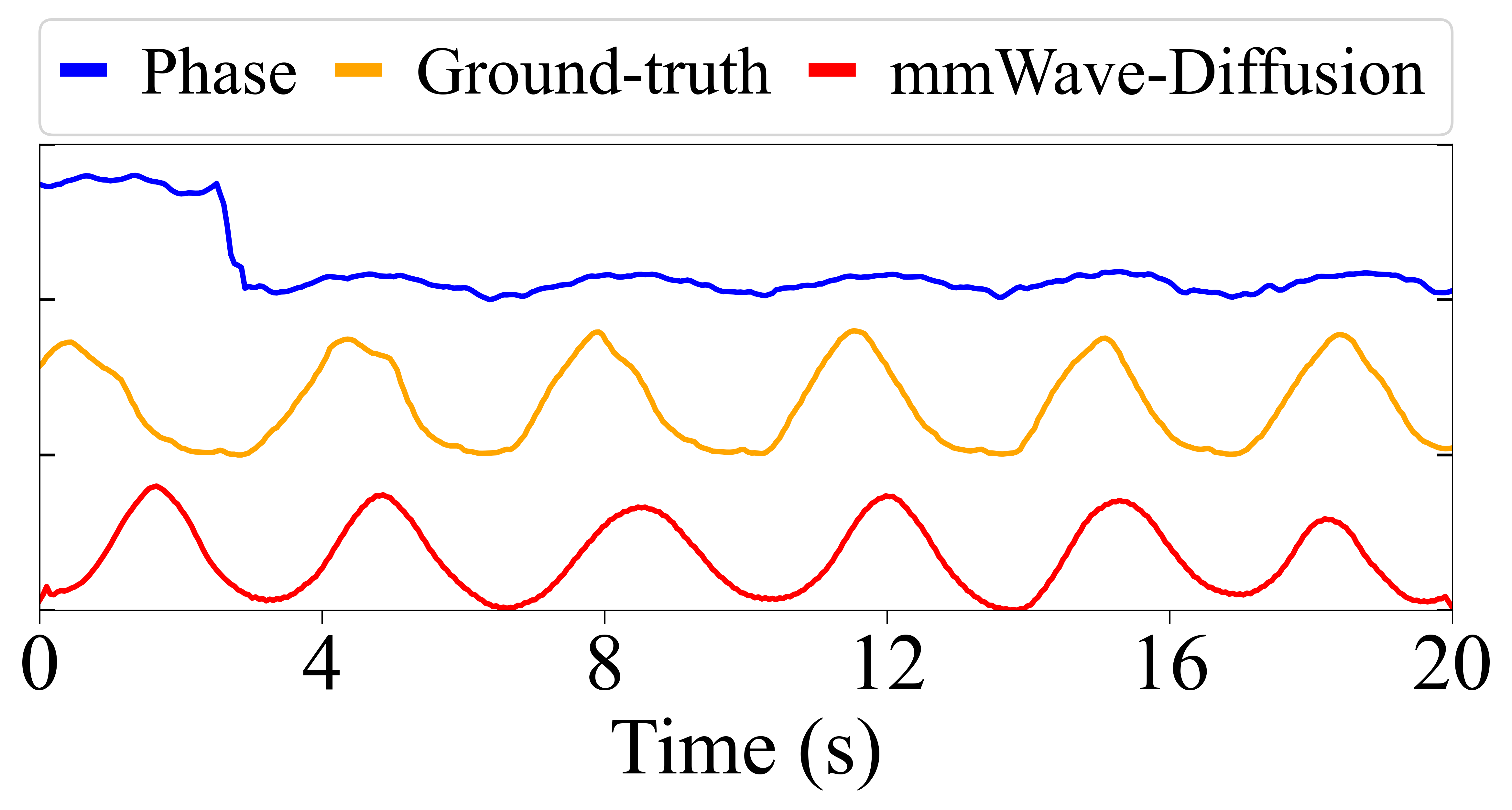}
        \caption{Page turning}
    \end{subfigure}
    \caption{Waveform reconstruction under diverse body micromotions.}
    \label{FIG:3}
\end{figure}

We further present qualitative evidence of mmWave-Diffusion’s micromotion-interference suppression. As shown in Fig.~\ref{FIG:3}, typical office micromotions (smartphone use, conversation, leg shaking, page turning) induce pronounced distortions in radar phase, markedly deviating from the near-sinusoidal respiratory ground truth. Within an observation-anchored diffusion paradigm, mmWave-Diffusion conditions on observations throughout forward and reverse steps to guide micromotion-interference removal; together with the tailored  RDT denoiser, artifacts are progressively suppressed and physiological structure restored. Consequently, the reconstructed waveforms closely match the ground truth in periodicity, morphology, and amplitude, demonstrating the robustness of our method.

\subsection{Ablation Study}
\label{ssec:Ablation}

We quantify the contribution of each RDT component through systematic ablations (Table~\ref{table:2}). Configurations are as follows: Variant1 removes multi-head self-attention; Variant2 removes the conditional phase tokens; Variant3 replaces banded-mask multi-head cross-attention with direct concatenation of the conditional and main tokens at the RDT input; Variant4 retains cross-attention but removes the banded mask. All variants underperform the full model, attesting to the substantive role of each component. Variant1 exhibits the largest degradation (CS/MAE = 0.753/1.274), indicating that high-fidelity reconstruction requires jointly learning a strong generative prior and explicitly modeling the intrinsic temporal dependencies of respiration. Variant2 likewise incurs a substantial degradation (0.758/1.062), underscoring that explicit phase conditioning reinforces physical alignment and strengthens conditional guidance. Variant4 (0.790/0.804) outperforms Variant3 (0.762/0.957) yet remains below the full model (0.811/0.631), showing that cross-attention effectively leverages conditional observations and that the banded mask enforces locality, mitigating spurious long-range correspondences and suppressing noise.

\subsection{Distance Robustness}
\label{ssec:Generalization}

\begin{figure}[t]
    \centering
    \begin{subfigure}[b]{0.49\linewidth}
        \centering
        \includegraphics[width=\linewidth]{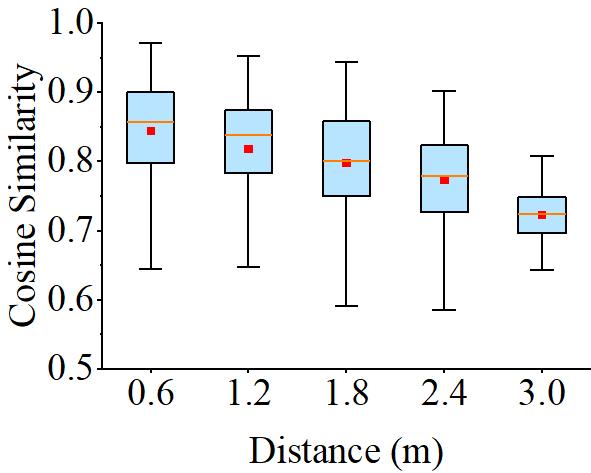}
        \caption{Cosine similarity}
    \end{subfigure}
    \hfill
    \begin{subfigure}[b]{0.49\linewidth}
        \centering
        \includegraphics[width=\linewidth]{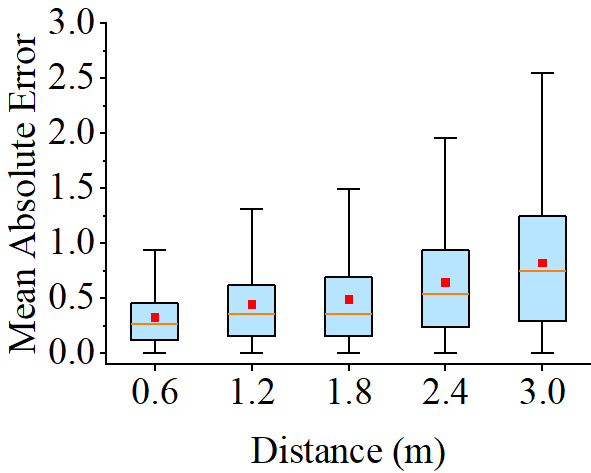}
        \caption{Mean absolute error}
    \end{subfigure}
    \caption{mmWave-Diffusion performance across sensing distances. (Red dots: the mean values; Orange solid lines: the medians.)}
    \label{FIG:4}
\end{figure}

Given that sensing distance is a principal determinant of Signal-to-Noise Ratio (SNR) in radar-based respiration monitoring, we collected five additional synchronous datasets spanning 0.6–3.0 m (15-minute sessions) to evaluate mmWave-Diffusion across sensing ranges. As summarized by the box plots in Fig.~\ref{FIG:4}, reconstruction accuracy declines approximately monotonically with increasing distance, consistent with propagation physics (greater range attenuates the echo and lowers SNR). Importantly, performance does not exhibit catastrophic degradation at 3.0 m: CS = 0.723 and MAE = 0.823 BPM. These results indicate that the proposed mmWave-Diffusion framework enables reliable respiratory monitoring under varying SNR conditions in real-world environments.

\begin{table}[t]
\caption{Radar Diffusion Transformer vs. Its Variants.}
\label{table:2}
\centering
\setlength{\tabcolsep}{4pt}     % 列间距
\scriptsize
\begin{tabular}{c c c c c c}
\toprule
\multirow{2}{*}{Method} &
\multicolumn{2}{c}{Waveform Reconstruction} &
\multicolumn{3}{c}{Frequency Estimation} \\
\cmidrule(lr){2-3} \cmidrule(lr){4-6}
& CS $\uparrow$ & MSE $\downarrow$ & MAE $\downarrow$ & RMSE $\downarrow$ & SD $\downarrow$ \\
\midrule
Variant1     & 0.753 & 0.118    & 1.274   & 1.599 & 1.435 \\
Variant2     & 0.758 & 0.102    &1.062    & 1.573 & 1.418 \\
Variant3     & 0.762 & 0.098    &0.957    & 1.376 & 1.261 \\
Variant4     & 0.790 & 0.096    &0.804    & 1.337 & 1.328 \\
\textbf{RDT}        & \textbf{0.811} & \textbf{0.079} & \textbf{0.631} & \textbf{1.175} & \textbf{1.035}  \\
\bottomrule
\end{tabular}
\end{table}

\section{Conclusion}
\label{sec:conclusion}

This paper proposes mmWave-Diffusion, an observation-anchored conditional diffusion framework for fine-grained mmWave radar respiratory monitoring under micromotion interference, effectively removing micromotion artifacts. Specifically, the forward process explicitly injects the residual carrying interference information along the path between the radar observation and the respiratory ground truth; the reverse process initializes within an observation-consistent neighborhood and explicitly conditions on the observation at each step to fully exploit it to constrain and guide the reconstruction. The accompanying Radar Diffusion Transformer (RDT) encodes physical constraints—temporal correspondence and local matching—as structural priors through patch-level positional alignment and banded-mask multi-head cross-attention, thereby propagating the observations throughout the denoising trajectory and enabling the targeted removal of micromotion-induced interference. Experimental results demonstrate that mmWave-Diffusion achieves state-of-the-art performance across multiple metrics for waveform reconstruction and respiratory-rate estimation. Ablation studies corroborate the necessity of each RDT component, and evaluations across sensing distances (0.6–3.0 m) demonstrate robustness under varying SNR conditions. Overall, mmWave-Diffusion sets a new benchmark for radar-based respiration monitoring and advances an effective paradigm for the deep integration of physical priors with conditional diffusion.

\vfill\pagebreak

\newpage
\bibliographystyle{IEEEbib}
\bibliography{refs}

\begin{thebibliography}{10}

\bibitem{Wang2024RF-GymCare}
Jianyang Wang, Dongheng Zhang, Binbin Zhang, Jinbo Chen, Yang Hu, and Yan Chen,
\newblock ``{RF-GymCare}: Introducing respiratory prior for {RF} sensing in gym
  environments,''
\newblock {\em Proceedings of the ACM on Interactive, Mobile, Wearable and
  Ubiquitous Technologies}, vol. 8, no. 3, pp. 1--28, 2024.

\bibitem{Zheng2021MoRe-Fi}
Tianyue Zheng, Zhe Chen, Shujie Zhang, Chao Cai, and Jun Luo,
\newblock ``{MoRe-Fi}: Motion-robust and fine-grained respiration monitoring
  via deep-learning {UWB} radar,''
\newblock in {\em Proceedings of the 19th ACM Conference on Embedded Networked
  Sensor Systems}, 2021, pp. 111--124.

\bibitem{Wang2024MM-FGRM}
Shuxuan Wang, Chong Han, Jian Guo, and Lijuan Sun,
\newblock ``{MM-FGRM}: Fine-grained respiratory monitoring using {MIMO}
  millimeter wave radar,''
\newblock {\em IEEE Transactions on Instrumentation and Measurement}, vol. 73,
  pp. 1--13, 2024.

\bibitem{Qiao2025Millimeter}
Xingshuai Qiao, Yaobin Su, Xiuping Li, and Tao Shan,
\newblock ``Millimeter-wave radar vital signs measurement with random body
  movement using missing data model,''
\newblock {\em IEEE Transactions on Instrumentation and Measurement}, vol. 74,
  pp. 1--14, 2025.

\bibitem{Wang2025GAWNet}
Yong Wang, Dongyu Liu, Chendong Xu, Bao Zhang, Yi~Lu, Kuiying Yin, Shuai Yao,
  and Qisong Wu,
\newblock ``{GAWNet}: A gated attention wavelet network for respiratory
  monitoring via millimeter-wave radar,''
\newblock {\em IEEE Signal Processing Letters}, vol. 32, pp. 3695--3699, 2025.

\bibitem{Zhang2023Pi}
Bo~Zhang, Boyu Jiang, Rong Zheng, Xiaoping Zhang, Jun Li, and Qiang Xu,
\newblock ``{Pi-ViMo}: Physiology-inspired robust vital sign monitoring using
  {mmWave} radars,''
\newblock {\em ACM Transactions on Internet of Things}, vol. 4, no. 2, pp.
  1--27, 2023.

\bibitem{Wang2026Fine}
Yong Wang, Chendong Xu, Bao Zhang, Zijun Huang, Dongyu Liu, Shuai Yao, Kuiying
  Yin, Qisong Wu, and Chaochao Wang,
\newblock ``Fine-grained contactless human respiratory measurement using
  millimeter-wave radar,''
\newblock {\em IEEE Transactions on Instrumentation and Measurement}, vol. 75,
  pp. 1--15, 2026.

\bibitem{Wu2024Contactless}
Yingxiao Wu, Haocheng Ni, Changlin Mao, and Jianping Han,
\newblock ``Contactless reconstruction of {ECG} and respiration signals with
  {mmWave} radar based on {RSSRnet},''
\newblock {\em IEEE Sensors Journal}, vol. 24, no. 5, pp. 6358--6368, 2024.

\bibitem{Bauder2025MM-MURE}
Chandler Bauder, Abdel-Kareem Moadi, Vijaysrinivas Rajagopal, Tianhao Wu, Jian
  Liu, and Aly~E. Fathy,
\newblock ``{MM-MURE}: {mmWave}-based multi-subject respiration monitoring via
  end-to-end deep learning,''
\newblock {\em IEEE Journal of Electromagnetics, RF and Microwaves in Medicine
  and Biology}, vol. 9, no. 1, pp. 49--61, 2025.

\bibitem{Zhang2025Diffusion}
Jinjin Zhang, Qiuyu Huang, Junjie Liu, Xiefan Guo, and Di~Huang,
\newblock ``Diffusion-4k: Ultra-high-resolution image synthesis with latent
  diffusion models,''
\newblock in {\em 2025 IEEE/CVF Conference on Computer Vision and Pattern
  Recognition (CVPR)}, 2025, pp. 23464--23473.

\bibitem{Rasul2021Autoregressive}
Kashif Rasul, Calvin Seward, Ingmar Schuster, and Roland Vollgraf,
\newblock ``Autoregressive denoising diffusion models for multivariate
  probabilistic time series forecasting,''
\newblock in {\em International Conference on Machine Learning}. PMLR, 2021,
  pp. 8857--8868.

\bibitem{Ho2020Denoising}
Jonathan Ho, Ajay Jain, and Pieter Abbeel,
\newblock ``Denoising diffusion probabilistic models,''
\newblock {\em Advances in Neural Information Processing Systems}, vol. 33, pp.
  6840--6851, 2020.

\bibitem{Dhariwal2021Diffusion}
Prafulla Dhariwal and Alexander Nichol,
\newblock ``Diffusion models beat gans on image synthesis,''
\newblock {\em Advances in Neural Information Processing Systems}, vol. 34, pp.
  8780--8794, 2021.

\bibitem{Peebles2023Scalable}
William Peebles and Saining Xie,
\newblock ``Scalable diffusion models with {Transformers},''
\newblock in {\em 2023 IEEE/CVF International Conference on Computer Vision
  (ICCV)}, 2023, pp. 4172--4182.

\bibitem{Diao2024Review}
Pape~Sanoussy Diao, Thierry Alves, Benoit Poussot, and Sylvain Azarian,
\newblock ``A review of radar detection fundamentals,''
\newblock {\em IEEE Aerospace and Electronic Systems Magazine}, vol. 39, no. 9,
  pp. 4--24, 2024.

\bibitem{Liaquat2024End}
Salman Liaquat, Nor~Muzlifah Mahyuddin, and Ijaz~Haider Naqvi,
\newblock ``An end-to-end modular framework for radar signal processing: A
  simulation-based tutorial,''
\newblock {\em IEEE Aerospace and Electronic Systems Magazine}, vol. 39, no. 9,
  pp. 98--118, 2024.

\bibitem{Yue2025Efficient}
Zongsheng Yue, Jianyi Wang, and Chen~Change Loy,
\newblock ``Efficient diffusion model for image restoration by residual
  shifting,''
\newblock {\em IEEE Transactions on Pattern Analysis and Machine Intelligence},
  vol. 47, no. 1, pp. 116--130, 2025.

\bibitem{Liu2024Residual}
Jiawei Liu, Qiang Wang, Huijie Fan, Yinong Wang, Yandong Tang, and Liangqiong
  Qu,
\newblock ``Residual denoising diffusion models,''
\newblock in {\em 2024 IEEE/CVF Conference on Computer Vision and Pattern
  Recognition (CVPR)}, 2024, pp. 2773--2783.

\bibitem{Shen2025Deep}
Qifan Shen, Xinwei Luo, and Long Chen,
\newblock ``A deep learning based iterative denoising algorithm for multiple
  frequency lines recovery,''
\newblock {\em Engineering Applications of Artificial Intelligence}, vol. 159,
  pp. 111601, 2025.

\bibitem{Song2020Denoising}
Jiaming Song, Chenlin Meng, and Stefano Ermon,
\newblock ``Denoising diffusion implicit models,''
\newblock {\em arXiv preprint arXiv:2010.02502}, 2020.

\bibitem{Wang2025SelaFD}
Yijun Wang, Yong Wang, Chendong Xu, Shuai Yao, and Qisong Wu,
\newblock ``{SelaFD}:seamless adaptation of vision {Transformer} fine-tuning
  for radar-based human activity recognition,''
\newblock in {\em 2025 IEEE International Conference on Acoustics, Speech and
  Signal Processing (ICASSP)}, 2025, pp. 1--5.

\bibitem{Wang2024Speech}
Yong Wang, Cheng Lu, Hailun Lian, Yan Zhao, Björn~W. Schuller, Yuan Zong, and
  Wenming Zheng,
\newblock ``Speech {Swin-Transformer}: Exploring a hierarchical {Transformer}
  with shifted windows for speech emotion recognition,''
\newblock in {\em 2024 IEEE International Conference on Acoustics, Speech and
  Signal Processing (ICASSP)}, 2024, pp. 11646--11650.

\bibitem{Lei2024Dataset}
Guangyu Lei, Wei Cheng, Xipeng Yin, and Yuqing Wu,
\newblock ``The dataset of multi-target vital signs monitored by {FMCW}
  radar,''
\newblock {\em Data in Brief}, vol. 57, pp. 111027, 2024.

\bibitem{Mauro2023Few-Shot}
Gianfranco Mauro, Maria De~Carlos~Diez, Julius Ott, Lorenzo Servadei, Manuel~P.
  Cuellar, and Diego~P. Morales-Santos,
\newblock ``Few-shot user-adaptable radar-based breath signal sensing,''
\newblock {\em Sensors}, vol. 23, no. 2, 2023.

\bibitem{Wang2025Hierarchical}
Yong Wang, Chendong Xu, Weirui Na, Dongyu Liu, Jiuqi Yan, Shuai Yao, and Qisong
  Wu,
\newblock ``Hierarchical {Transformer} with auxiliary learning for
  subject-independent respiration emotion recognition,''
\newblock {\em IEEE Sensors Journal}, vol. 25, no. 16, pp. 31290--31301, 2025.

\bibitem{Islam2020NonContact}
Shekh M~M Islam, Naoyuki Motoyama, Sergio Pacheco, and Victor~M. Lubecke,
\newblock ``Non-contact vital signs monitoring for multiple subjects using a
  millimeter wave {FMCW} automotive radar,''
\newblock in {\em 2020 IEEE/MTT-S International Microwave Symposium (IMS)},
  2020, pp. 783--786.

\end{thebibliography}

\end{document}